\documentclass{WileyASNA-v1} 
\usepackage{graphicx}
\usepackage{amsmath}
\usepackage{amsfonts}
\bibpunct{(}{)}{;}{a}{}{,} 
\usepackage{fancyhdr}
\usepackage{orcidlink}

\articletype{Article}%

\received{\today}
\revised{...}
\accepted{...}
\begin{document}

\title{On the age distribution of Classical Cepheids in the Galaxy}

\author[1,2,3]{Friedrich Anders$^{*,}$}
\author[1,2,3]{Chloé Padois}
\author[1]{Marc Vilanova Sar}
\author[1,2,3]{Marcin Semczuk}
\author[1,2,3]{Marc del Alcázar-Julià}
\author[1,2,3]{Francesca Figueras}

\authormark{Anders, Padois, \textsc{et al}: Age distribution of Classical Cepheids in the Galaxy}

\address[1]{\orgdiv{Departament de Física Quàntica i Astrofísica (FQA)}, \orgname{Universitat de Barcelona (UB)}, \orgaddress{\state{C Martí i Franqués, 1, 08028 Barcelona}, \country{Spain}}}
\address[2]{\orgdiv{Institut de Ciències del Cosmos (ICCUB)}, \orgname{Universitat de Barcelona (UB)}, \orgaddress{\state{C Martí i Franqués, 1, 08028 Barcelona}, \country{Spain}}}
\address[3]{\orgname{Institut d'Estudis Espacials de Catalunya (IEEC)}, \orgaddress{\state{Edifici RDIT, Campus UPC, 08860 Castelldefels (Barcelona)}, \country{Spain}}}
\corres{*\email{fanders@fqa.ub.edu}}

\abstract{We revisit the problem of the positive correlation between age and Galactocentric distance seen in Galactic Classical Cepheids, which at first sight may seem counter-intuitive in the context of inside-out galaxy formation. 
To explain it, we use the Besançon Galaxy Model and a simulation of star particles in the Galactic disc coupled with stellar evolutionary models. We then select Classical Cepheids from this simulation and test in qualitative terms which ingredients are necessary to find agreement with the observational data. We show that 
the interplay of the Galactic disc's metallicity gradient and the metallicity dependence of the Cepheids' life-time in the instability strip results in a pronounced positive age-Galactocentric distance relation. This renders a reconstruction of the recent star-formation history based on Classical Cepheids unrealistic. It also has important consequences on our interpretation of the observed scatter about the radial metallicity gradient measured with Galactic Classical Cepheids.}

\keywords{Cepheids, stars: variables: general, Galaxy: structure, stars: fundamental parameters, Galaxy: disk, Galaxy: evolution}

\jnlcitation{\cname{%
\author{Anders, F.}, 
\author{Padois, C.}, 
\author{Vilanova Sar, M.}, 
\author{Semczuk, M.}, 
\author{del Alcázar-Julià, M.}, and 
\author{Figueras, F.}} (\cyear{2025}), 
\ctitle{On the age distribution of Classical Cepheids in the Galaxy}, \cjournal{AN}, \cvol{in press}.}

\maketitle

\section{Introduction}\label{sec:intro}

In the past ten years, classical Cepheid variables (CCs) have attracted more and more attention from the Galactic astrophysics community due to their potential for creating detailed large-scale maps of the Milky Way \citep[e.g.][]{Bahner1962, Kraft1963, Feast2014, Lemasle2018, Skowron2019, Chen2019, Dekany2019, Lemasle2022, GaiaCollaboration2023, Drimmel2023, Semczuk2023}. CCs are evolved intermediate-mass ($3M_{\odot}-10M_{\odot}$; e.g. \citealt{Gieren1989, Georgy2013}) pulsating stars inhabiting the instability strip (IS) of the Hertzsprung-Russell diagram \citep{Gautschy1966} that have served as reliable Galactic and extragalactic distance indicators for more than one hundred years \citep{Leavitt1912, Lemaitre1927, Hubble1929, Anderson2024, Skowron2024}. 

The Milky Way's CCs are population I stars with regular periods typically below 10 days (although periods of up to 78 days have been detected; \citealt{Soszynski2024}), disc-like metallicities ($-0.4\lesssim$ [Fe/H] $\lesssim +0.4$; e.g. \citealt{Genovali2014}) and effective temperatures ranging from $\sim4500$ K to $\sim7000$ K \citep{Lemasle2020}. 
CCs are interesting tracers of young stellar populations in the Galaxy and beyond for the following reasons: 1. They are young (typically between 20 and 200 Myr) and their abundance and luminosity allows us to trace a significant fraction of the full Galactic CC population \citep{Pietrukowicz2021}, and 2. They follow a tight period-age relation (see \citealt{Efremov1977, Bono2005, Anderson2016, DeSomma2021}), which enables us to determine their age with better precision and accuracy compared to other young tracers. 

CCs have several applications due to their period-luminosity relation \citep{Leavitt1912}, which provides more precise distance estimates at large distances (several kpc) than the parallax method with current {\it Gaia} data \citep{GaiaCollaboration2023}. In the case of the Milky Way, properties like the Galactic rotation curve \citep{Mroz2019}, chemical-abundance gradients \citep[e.g.][]{Lemasle2007, Genovali2014, Genovali2015, Luck2018, daSilva2022, Kovtyukh2022}, the shape and evolution of the Galactic warp \citep{Chen2019, Dekany2019, Skowron2019AcA, Lemasle2022, Dehnen2023, Cabrera-Gadea2024}, spiral-arm configuration \citep[e.g.][]{Skowron2019, Poggio2021, GaiaCollaboration2023, Semczuk2023}, or the dark-matter content \citep[e.g.][]{Ablimit2020} can been inferred based on CC distance measurements. For nearby galaxies, most notably the Magellanic Clouds, CCs can be used to study their three-dimensional structure \citep{Scowcroft2016, Jacyszyn-Dobrzeniecka2016, Ripepi2017}. In particular, the age and distance of the Magellanic bridge has been measured with CC measurements \citep{Jacyszyn-Dobrzeniecka2020}. Last not least, CCs are the primary standard candles that fix the cosmic distance scales \citep[e.g.][]{Riess2012, Inno2013, Breuval2024, CruzReyes2023, Anderson2024}. For a recent review on CCs and other Cepheid-type variables we refer to \citet{Bono2024}.

\begin{figure}
    \centering
    \includegraphics[width=0.5\textwidth]{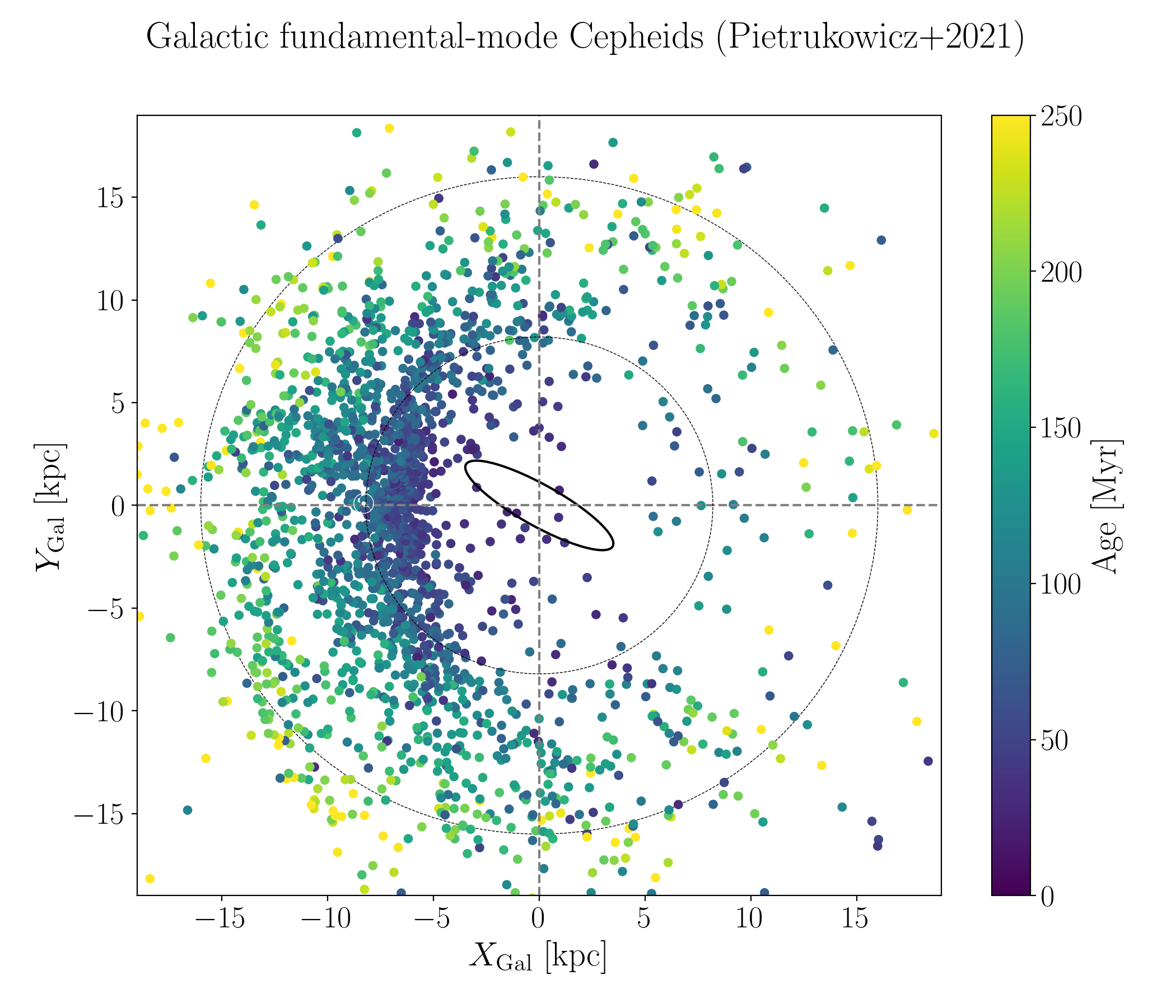}
    \includegraphics[width=0.5\textwidth]{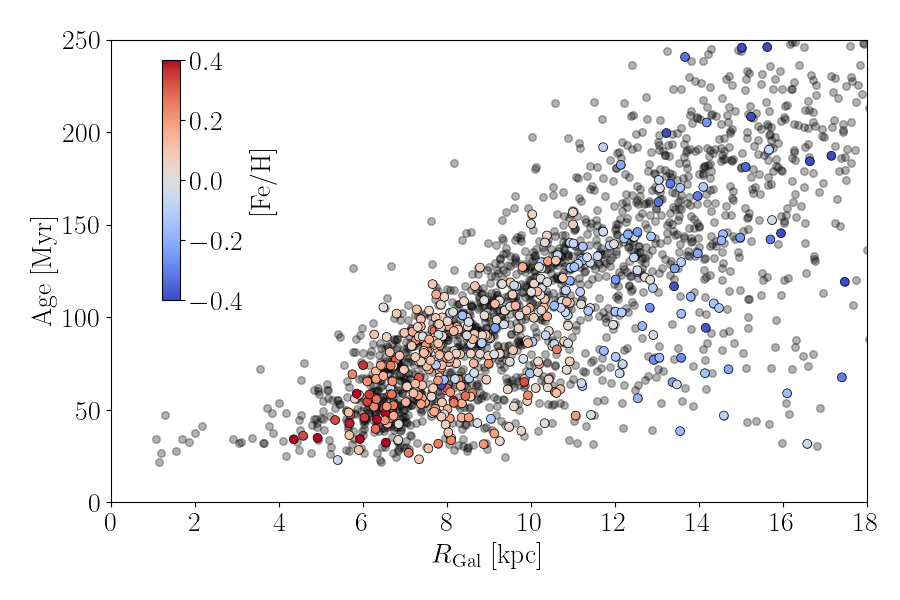}
    \caption{Distribution of known fundamental-mode classical Cepheids \citep{Pietrukowicz2021} in the Galaxy. Top panel: in Galactocentric Cartesian coordinates, colour-coded by age. Bottom panel: Age vs. Galactocentric distance, colour-coded by metallicity (for the objects with spectroscopic [Fe/H] measurements compiled by \citealt{Ripepi2022}).}
    \label{fig:xy_age}
\end{figure}

A particularly interesting and puzzling observable in the Galactic astrophysics context -- and the topic of this short paper -- is the age distribution of Milky Way Cepheids as a function of Galactocentric distance (see Fig. \ref{fig:xy_age}). All recent studies of the large-scale distribution of Galactic CCs, independent of the calibration of the period-age relation, have found a clear positive age gradient with Galactocentric distance \citep[e.g.][]{Dekany2019, Skowron2019, DeSomma2020, Pietrukowicz2021}. For example, \citet{DeSomma2020} suggested that Cepheids could potentially be used to perform an age tomography of the Galaxy that would allow us to trace the recent star-formation history of the Milky Way. In the Magellanic Clouds, a similar behaviour was observed by \citet{Joshi2019}. 
At first sight, the observed trend may indeed seem counter-intuitive, because our Galaxy is widely accepted to have undergone an inside-out formation process, meaning that star-formation progressively reached the outer regions of the Galaxy over the past billion years (e.g. \citealt{Lacey1985, Matteucci1989, Ferrini1994, Chiappini2001, SanchezBlazquez2011, Matteucci2012, Frankel2019, Queiroz2020, Spitoni2021, Prantzos2023, Chen2025}). From a purely Galactic point of view, one might therefore be tempted to invoke a drastically changing star-formation history on short timescales, or even a change in the initial-mass function to explain this phenomenon. Evidence from star and star-cluster counts suggests that the extended solar vicinity (i.e. the nearest $\sim 2-3$ kpc) might indeed have experienced a recent star burst (around $\sim 6-30$ Myr ago; \citealt{Morales2013, Anders2021, Zari2023}, but see also \citealt{Soler2023, Gallart2024, delAlcazar-Julia2025}). This age interval, however, is outside the age range well traced by Galactic CCs.

In this paper, we show that most of the signal contained in the age vs. Galactocentric distance ($R_{\rm Gal}$) distribution of Cepheids can be explained by the negative radial metallicity gradient in the Galactic thin disc, coupled with the metallicity dependence of the instability strip \citep[e.g.][]{DeSomma2021}. This idea was previously explored in the supplementary material of \citet{Skowron2019} using simple simulations. In this paper we approach the problem in two complementary ways: 1. by using the well-tested Besançon Galaxy Model (BGM), and 2. with custom simple but flexible simulations using BaSTI-IAC stellar models. 

The paper is structured as follows: In Sect. \ref{sec:data} we present the observational data used for this study, and in Sect. \ref{sec:besancon} we show that the observed age vs. $R_{\rm Gal}$ trend in CCs is qualitatively reproduced with a stellar-population synthesis model (using the  Besançon Galaxy Model presented in \citealt{Mor2019}). In Sect. \ref{sec:simu} we present a set of simple but tailored simulations in which we have full control over the different model ingredients of the simulated CC population to study the effects of the different choices (of e.g. the definition of the instability strip or the Galactic radial metallicity gradient). We conclude the paper with a brief discussion in Sect. \ref{sec:conclusions}.

\section{Observational data}\label{sec:data}

\begin{figure*}
\begin{center}
\includegraphics[width=0.95\textwidth]{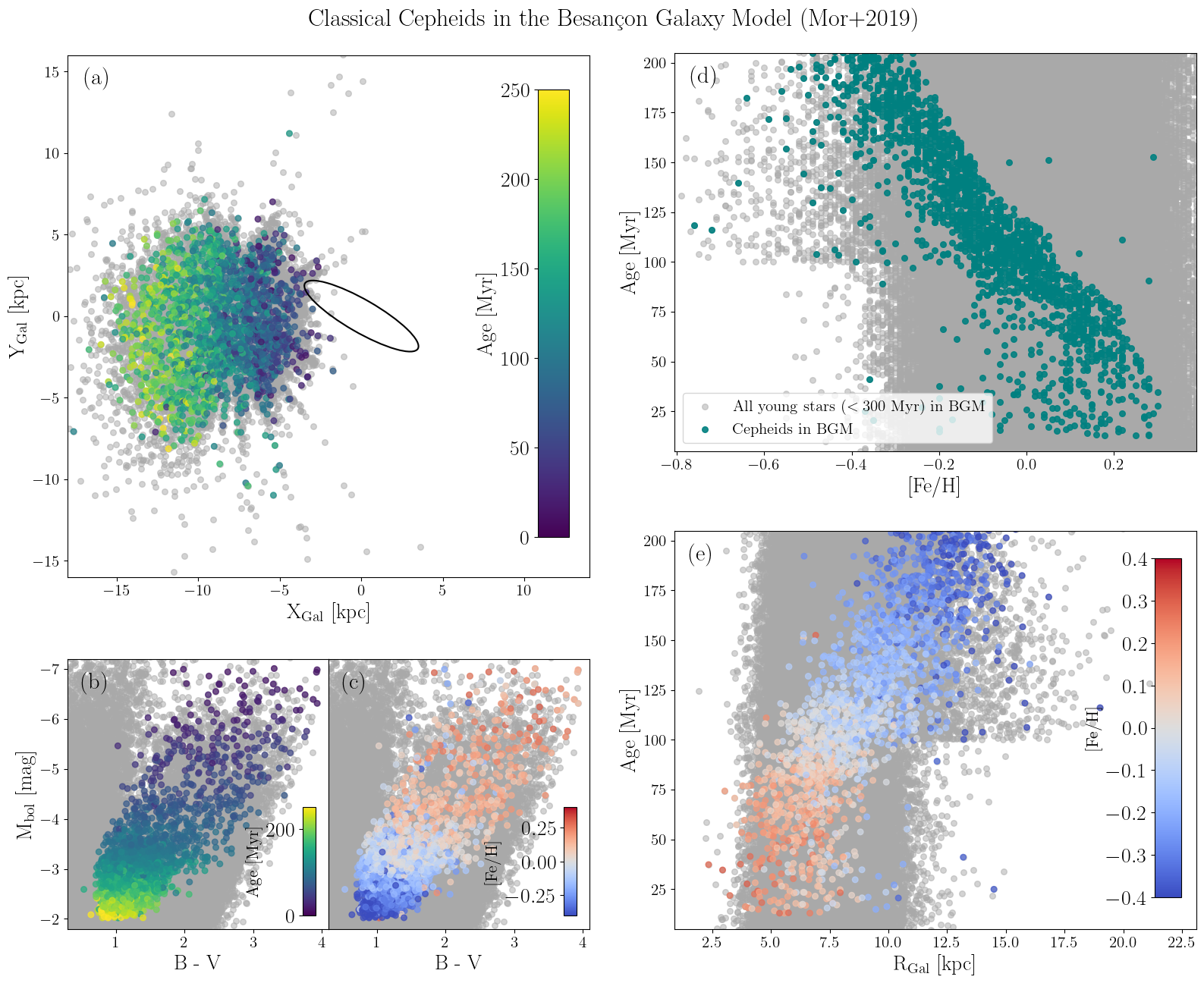}
\caption{Young stars ($<300$ Myr) in the extended solar-vicinity ($G<13$) BGM simulation of \citet{Mor2019}. In each of the panels, the coloured points correspond to CCs, while the grey dots in the background correspond to all simulated stars. Top left panel (a): Distribution in the Galactic plane, similar to the upper panel of Fig. \ref{fig:xy_age}. Lower-left panels (b) and (c): selection of CCs in the CMD, coloured by age and [Fe/H]. Upper right panel (d): Age vs. [Fe/H]. Lower right panel (e):  Age vs. $R_{\rm Gal}$, as in the lower panel of Fig. \ref{fig:xy_age}.
\label{fig:besancon}
}
\end{center}
\end{figure*}

We start from the catalogue of \citet{Pietrukowicz2021}, which is the largest, cleanest, and most up-to-date catalogue of Galactic CCs available\footnote{{\tt \url{https://www.astrouw.edu.pl/ogle/ogle4/OCVS/allGalCep.listID}}, April 2024 version}. It gathers curated observational data for Galactic CCs from many different sources, such as the Optical Gravitational Lensing Experiment (OGLE; \citealt{Udalski1997, Udalski2008, Udalski2015}), ASAS \citep{Pojmanski1997}, ASAS-SN \citep{Kochanek2017}, ATLAS \citep{Tonry2018, Heinze2018}, {\it Gaia} \citep{GaiaCollaboration2016, GaiaCollaboration2021}, NSVS \citep{Hoffman2009}, VVV \citep{Minniti2010}, WISE \citep{Chen2018}, and ZTF \citep{Bellm2019}. The April 2024 version of the list contains 3647 Cepheids, of which 3642 have astrometry and photometry from {\it Gaia} DR3 \citep{GaiaCollaboration2023}. From this catalogue, we selected the 2247 fundamental-mode CCs, for which we calculated their distances following \citet{Ripepi2022} and their ages using the recipe of \citet{Anderson2016}, assuming the Galactic radial abundance gradient of \citet{Genovali2014} for CCs without a measured iron abundance in the compilation \citet{Ripepi2022}. We can see in Fig. \ref{fig:xy_age} how the stars from the catalogue are distributed across the XY Galactic plane, colour-coded by their age (see Sect. \ref{sec:intro}).

Regarding the age scale, we tested various choices of the period-age relation (in-/excluding metallicity terms as in \citealt{Dekany2019}), and found no qualitative change of the observed age trend with Galactocentric distance (in agreement with e.g. \citealt{DeSomma2020}). In fact, the observed age-$R_{\rm Gal}$ distribution in Fig. \ref{fig:xy_age} cannot arise from an erroneous period-age relation: when plotting the observed periods instead of the inferred ages, we encounter longer periods towards the Galactic centre and shorter ones towards the Galactic outskirts, which implies that the trend persists regardless of the details of the period-age relation.

\section{Besançon Galaxy model simulation}\label{sec:besancon}

In this section, we verify that the observed positive trend in the age vs. $R_{\rm Gal}$ diagram is reproduced by a well-tested stellar population-synthesis model. To this end, we employ the magnitude-limited ($G<13$) Besançon Galaxy Model (BGM) simulation presented in \citet{Mor2019}. Prior to that publication, the BGM \citep{Robin2003, Czekaj2014, Robin2022} had already been successfully used to constrain the high-mass slope of the initial mass function in the Galactic disc using CCs \citep{Mor2017}, among many other science cases \citep[e.g.][]{Robin2014, Amores2017, Lagarde2017}.

In the BGM version used in this simulation \citep{Mor2019}, the young disc populations follow an exponential disc profile with a pronounced radial metallicity gradient, a doubly-broken power-law initial-mass function that does not vary with time or metallicity, and a star-formation rate that is piecewise constant (i.e. is slightly varying for $<100$ Myr and $100-1000$ Myr). The stellar evolutionary models used by the BGM in the stellar mass range relevant for CCs are the non-rotating scaled-solar Padova models from \citet{Bertelli2009}. We refer to their Sect. 6 for an extensive discussion and comparison of these models with Cepheid observations and other models available at the time.
To select CC variables in the BGM, \citet{Mor2017} defined the edges of the instability strip (IS) in the colour-magnitude diagram following \citet{Bono2000} for the blue edge and \citet{Fiorentino2013} for the red edge, which results in a reasonable selection of CC stars. In the simulation (and in the \citealt{Bertelli2009} models in general), the dependence of the boundaries of the IS on metallicity is negligible, so that the definition of the instability strip employed by \citet{Mor2017} does not depend on metallicity.

Figure \ref{fig:besancon} summarises our findings from the BGM simulation. The figure focuses on the young local disc population (age $<300$ Myr; grey points in the background of each panel), and shows how the CCs were selected based on cuts in the colour-magnitude diagram (panels (b) and (c)). When comparing Fig. \ref{fig:besancon} panels (a) and (e) with the data shown in Fig. \ref{fig:xy_age}, we find that the simulation of \citet{Mor2019} reproduces the main trend seen in the data: younger Cepheids are tendentially found towards the inner disc, while older ones reside in the outer disc. The very good qualitative agreement between the data and the model arises despite the fact that the model was not optimised to account for the selection function of the CC sample. 

Figure \ref{fig:besancon} demonstrates that the resulting stellar age distribution seen in panel (e) is largely a consequence of the tight correspondence of age and metallicity in the IS region inhabited by the CCs (panels b and c), which is not present in the underlying population of young stars (panel d). Because of the higher metallicities in the inner Galactic disc, we preferentially find higher-mass, longer-period Cepheids towards there. In the outer disc, which has sub-solar metallicity, we tend to find lower-mass, shorter-period CCs.

\section{Custom simulations using BASTI models}\label{sec:simu}

To study the problem in a fully controlled setting, we create a set of tailored simulations that attempt to model the radial age distribution of Galactic CCs (see Fig. \ref{fig:xy_age}), testing different assumptions. This simulation consists of three steps. First, we simulate star particles following several basic properties of the Galactic disc population (density profile, star-formation rate, initial mass function). Then we associate each simulated star with a stellar evolutionary model. Finally, we define the limits of the IS to select and study the CCs that result from the different modeling assumptions, comparing them to the observational data. 

\subsection{Simulation setup}\label{sec:simu_1}

\begin{figure}
    \centering
    \includegraphics[width=0.5\textwidth]{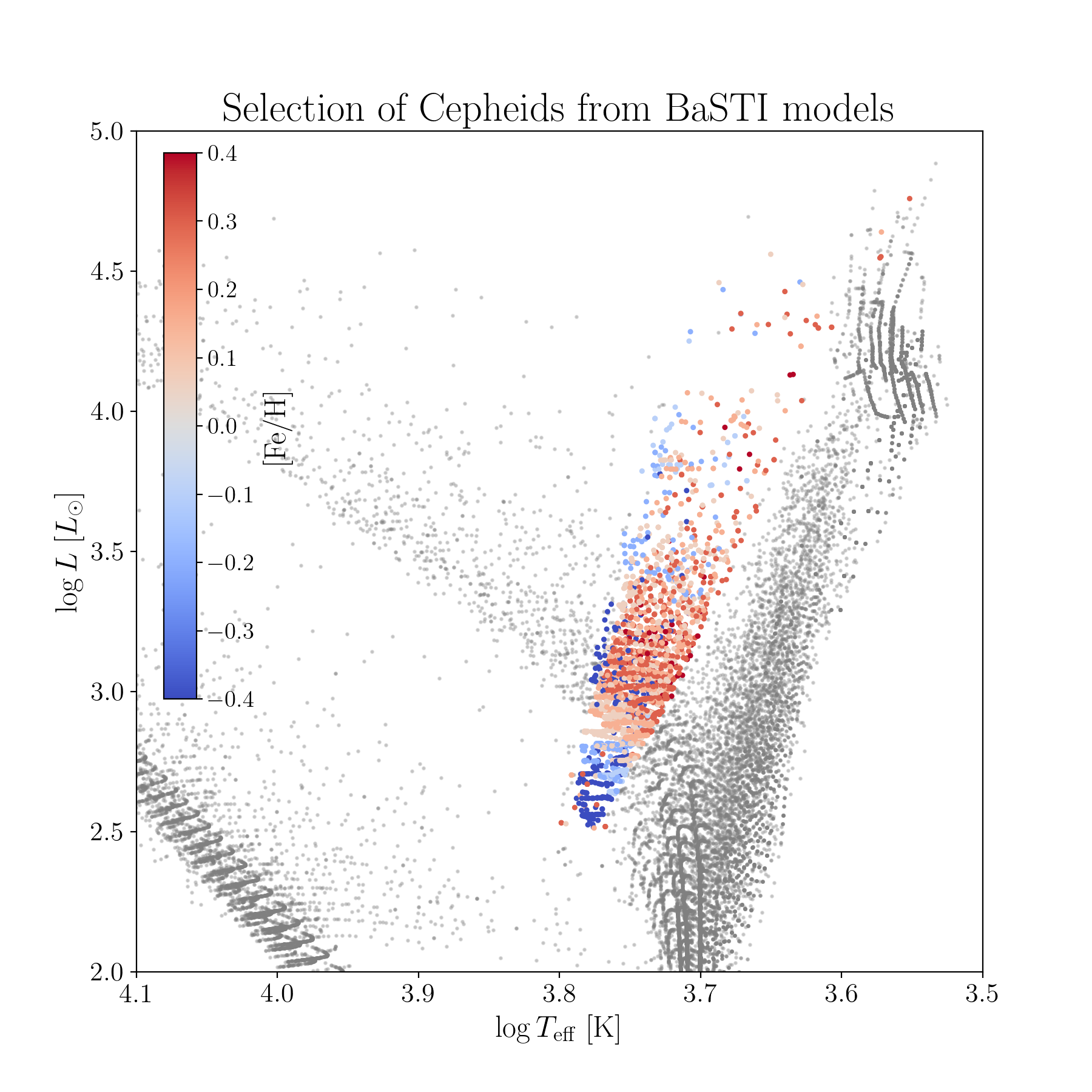}
    \caption{Simulated population of young stars, using the BaSTI evolutionary models \citep{Hidalgo2018}. The Cepheids, selected from the interpolated IS definition of \citet{DeSomma2021}, are colour-coded by [Fe/H].}
    \label{fig:hrd}
\end{figure}

We create a sample of $10^7$ disc star particles using the following initial conditions: an exponentially decreasing density profile with scale length $h_R=3.5$ kpc, a SFR constant in time (we are only simulating the past 500 Myr) and a non-varying Salpeter-like initial mass function in the range of $3M_\odot-15M_\odot$ ($\alpha=2.35$; \citealt{Salpeter1955}).
In order to study how the stars behave in terms of metallicity, we assign metallicities to each generated star following the radial metallicity gradient (modulo a 0.05 dex Gaussian scatter) measured by \citet{Genovali2014} using high-resolution spectroscopy of 450 CCs covering the Galactocentric distance range of 5-19 kpc. To cover a similar extent of the Galactic disc, in our simulation we only consider Galactocentric distances between 4 and 18 kpc. Moreover, for the purpose of filtering which synthetic stars may correspond to a CC, a loose age cut has been set (using the fact that these stars have already left the main sequence): only star particles with ages greater than 90\% of their lifetime can potentially be in the CC evolutionary phase (e.g. \citealt{Kippenhahn2013}, Chapter 31). Applying these filters results in 247 thousand evolved stars in the initial sample.

In the next step, each of the potential CC stars was associated with a stellar model from the BaSTI updated library of stellar evolutionary models\footnote{{\tt \url{http://basti-iac.oa-abruzzo.inaf.it/tracks.html}}} \citep{Hidalgo2018}, by default with solar-scaled abundance mixture and without overshooting, mass loss, or diffusion. We downloaded a grid of 2.2 million models with masses between 3 $\rm{M_{\odot}}$ and 15 $\rm{M_{\odot}}$ (with a resolution of 0.1 $\rm{M_{\odot}}$) and metallicities [Fe/H] $\in \{-0.9, -0.6, -0.4, -0.2, -0.1, +0.06, +0.15, +0.3, +0.45\}$. Then we matched each star with the closest evolutionary model by minimizing the Euclidean distance in the $\{{\rm \log age, mass, [Fe/H]}\}$ space.

The next step was to determine which of the possible simulated stars are found in the IS, and therefore can be defined as CCs. The correct definition of the IS is still an active field of debate (e.g. \citealt{Deka2024}), so we test the impact of different IS limits on our results. By default, we use IS limits of \citet{DeSomma2021} that were derived using the same stellar evolutionary models that we use in our toy population synthesis model based on the BaSTI grid. 

 \begin{figure*}
    \centering
    \includegraphics[width=0.9\textwidth]{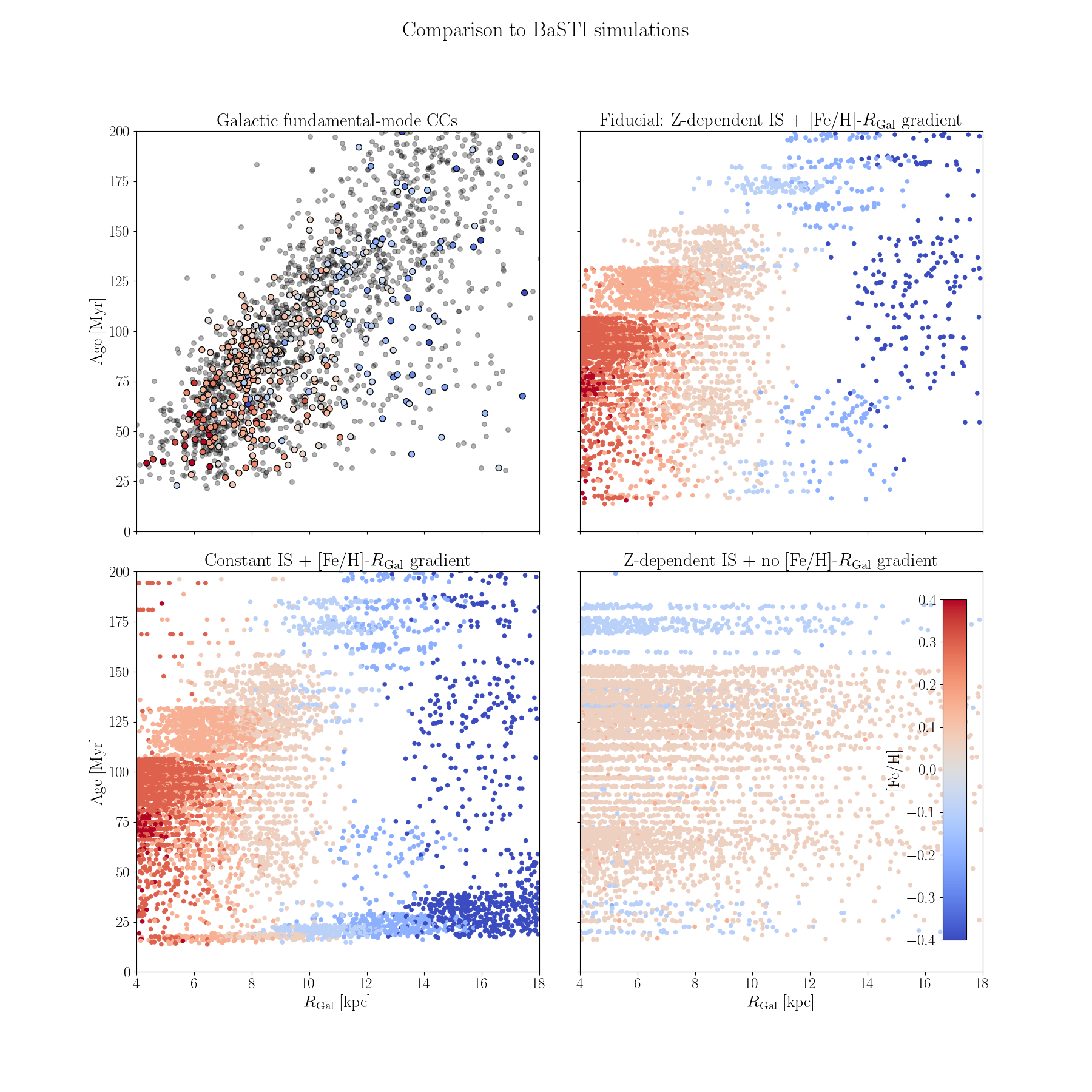}
    \caption{Age vs Galactocentric radius for observed and simulated CCs colour-coded by metallicity. Top left: Observational data from \protect\cite{Pietrukowicz2021}, colour-coded by [Fe/H] (when available) as in Fig. \ref{fig:xy_age}. Top right: Simulated Cepheids with metallicity-dependent IS and Galactic [Fe/H] gradient applied. Bottom left: Simulated Cepheids with constant IS and [Fe/H] gradient applied. Bottom right: Simulated Cepheids with metallicity-dependent IS and no [Fe/H] gradient applied.}  
    \label{fig:age_radius}
\end{figure*}

Some studies have modeled the red and blue edges of the IS as constant lines in the Hertzsprung-Russell or {\it Kiel} diagram (see Sect. \ref{sec:besancon}), while more recent stellar models and studies suggest a more pronounced metallicity dependence, since observational data show that with higher metallicity the IS becomes redder (see e.g. \citealt{Cox1980, DeSomma2021, DeSomma2022, Espinoza-Arancibia2024}). This phenomenon can be explained by two reasons: on the one hand, as metallicity increases, hydrogen is less abundant, which delays the beginning of pulsation to lower effective temperatures; on the other hand, the increased contribution of iron delays the pulsation to the red edge due to convection. This dependence has been studied and quantified in recent years \citep[e.g.][]{Anderson2016, DeSomma2021, DeSomma2022}.

To test the effect of this $Z$-dependence of the IS on our synthetic population, we model the IS limits both with a dependence on metallicity and without it. In the first case, we use the work of \citet{DeSomma2021} in which the IS limits are determined for three bins in metallicity: Small Magellanic Cloud ($Z=0.004, Y=0.25$; their Table 2), Large Magellanic Cloud ($Z=0.008, Y=0.25$; their Table 3), and M31 ($Z=0.03, Y=0.28$; their Table 4). By linearly interpolating between these values, we obtain a continuous relation of the blue and red edges of the IS as a function of metallicity. Since the limits defined by \cite{DeSomma2021} depend on the absolute metallicity and we want to express our results in terms of the iron-to-hydrogen abundance ratio [Fe/H], we use the canonic transformation [Fe/H]$=10^{Z-Z_\odot}$. In the second case (no $Z$ dependence of the IS), we use the definition given by \citet[][used in Sect. \ref{sec:besancon}]{Mor2017}, who applied the red edge described by \citet{Bono2000} and the blue edge by \citet{Fiorentino2013}.

In Fig. \ref{fig:hrd} we show the Hertzsprung-Russell diagram of the simulated stars classified as possible CCs, highlighting the stars that actually fall inside the IS, colour-coded by metallicity, for the case in which the IS limits are metallicity-dependent and the \cite{Genovali2014} gradient is applied. The CC region contains 2165 stars.
As expected, the simulated CCs have effective temperatures from 3000K to 6000K and luminosities ranging from $\log(L/L_{\odot})=2.5$ up to 5. As in the BGM simulation presented in Sect. \ref{sec:besancon}, we tend to find more metal-poor stars at lower luminosities.

\subsection{Analysis}\label{sec:analysis}

In this section we study the simulated CCs (testing some of the underlying assumptions) by comparing their radial age distribution to the observational data.  Figure \ref{fig:age_radius} shows the main result of this paper: it shows that, similar to our findings using the Besançon simulation in Sect. \ref{sec:besancon}, the Galactic radial metallicity gradient is the main ingredient that shapes the observed age vs. $R_{\rm Gal}$ distribution. It also shows that a metallicity-dependent IS helps to obtain a better match with the observed CC population.

In the top right panel of Fig. \ref{fig:age_radius}, we show the results for the age vs. $R_{\rm Gal}$ distribution of our fiducial BaSTI simulation, including both the radial metallicity gradient and the $Z$-dependent IS. The resulting radial age distribution of the 2\,165 CCs follows a clear positive trend, and the density overall agrees well with the observational data seen in the top left panel of Fig. \ref{fig:age_radius}. 

In the case of the simulation with the Galactic abundance gradient but constant IS (Fig. \ref{fig:age_radius}, lower left panel), 3739 stars are found inside the IS and the match is worse than in the fiducial case. In particular, there is a much larger number of young CCs, including at Galactocentric distances higher than 15 kpc, which is not observed in the Milky Way. This can be explained by the wider IS limits that were used in this case, so that too many simulated stars which are not actually CCs are found in the IS. 

Finally, the bottom right panel of Fig. \ref{fig:age_radius} shows the result of the BaSTI simulation when no metallicity gradient, only the $Z$-dependent IS, is applied, resulting in 3139 stars that do not reproduce the observed trend in the age vs. $R_{\rm Gal}$ distribution at all.

\begin{figure}
    \centering
    \includegraphics[width=0.5\textwidth]{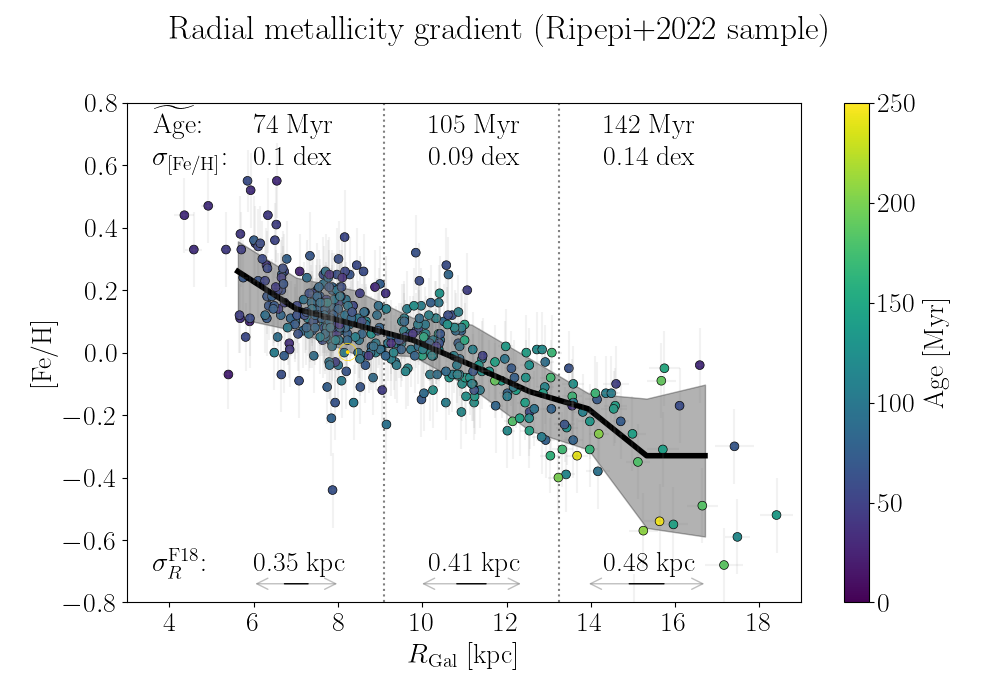}
    \caption{Radial metallicity profile ([Fe/H] vs. $R_{\rm Gal}$) of the spectroscopic sample of fundamental-mode CCs compiled by \citet{Ripepi2022}. Only fundamental-mode CCs are plotted; the vertical errorbars correspond to the values given in the catalogue; the horizontal errorbars assume a conservative distance precision of 5\%. The values displayed in the top of the plot show the median age, and the abundance dispersion around the main trend for the three $R_{\rm Gal}$ ranges highlighted by the vertical lines. The values at the bottom of the plot correspond to the radial migration scale corresponding to these ages assuming the simple diffusion model of \citet{Frankel2018}.}
    \label{fig:gradient}
\end{figure}

\section{Discussion and conclusions}\label{sec:conclusions}

In this short paper, we have revisited the age vs. Galactocentric distance distribution observed in Galactic CC variables. We find that the observed positive correlation of age and $R_{\rm Gal}$ is largely a consequence of the negative radial metallicity gradient in the Milky Way disc, coupled with the metallicity-dependent time interval that Cepheids spend inside the instability strip. The dependence of the IS on metallicity is an important secondary effect that should be taken into account when using last-generation stellar models (see Fig. \ref{fig:age_radius}). At the very least, the definition of the IS needs to be consistent with the evolutionary models used to select CC stars. In the case of the BaSTI stellar models, we find that the IS has to have a metallicity dependence similar to \citet{DeSomma2021, DeSomma2022} in order to agree with the observational trend. We stress, however, that the exact definition of the IS edges does not drastically influence the observed trends, as highlighted by the BGM simulation studied in Sect. \ref{sec:besancon}, which used a different set of evolutionary models and constant IS red and blue edges. \citet{Deka2024} have also recently found using the radial stellar pulsation models \citep{Paxton2019} available in the MESA code \citep{Paxton2011} that 90\% of the observed CCs fall within their predicted blue and red IS edges, regardless of their considered variations in the convection physics. 

Despite inside-out formation of the Mily Way disc, as evidenced by its negative radial metallicity gradient, older CCs accumulate in the outer disc precisely as a consequence of this radial metallicity gradient. The atmospheric metallicity modifies both the time spent on the instability strip and the range of masses for which a star will actually enter the instability strip.
This natural explanation was already put forward by \citet{Skowron2019}: {\it "The spatial distribution of ages shows that the closer to the Galactic center, the younger Cepheids we observe. The age distribution of Cepheids does not directly reflect that of all stars present in the disk. For example, the absence of short-period Cepheids in the inner disk does not necessarily mean that older (less massive) stars are absent in that region, as the higher metallicity there would not have produced Cepheids in the older population."}
Our findings (based on different stellar models and using two complementary methodologies) are in agreement with those of \citet{Skowron2019}; we expand on their results by showing that indeed both the negative radial metallicity gradient {\it and} the metallicity-dependent IS are necessary to be able to explain the trend seen in the observations.

The main consequence of this effect in the Galactic evolution context is that Cepheids should not be regarded as unbiased tracers of the young disc population, neither in the Milky Way nor in any other Galaxy that displays significant metallicity variations.
In other words, when using CCs as tracers of age trends in a galaxy, the effects of stellar evolution may greatly outweigh the effects of galaxy evolution. In particular, without detailed forward modeling of all involved stellar evolution effects, CCs should not be used to trace the recent star-formation history of the Milky Way disc. Of course, these words of caution are less drastic when CCs are considered as a monolithic "young" population (compared to the typical time scales of Galactic evolution; $\gtrsim 500$ Myr). But even in the case of the interpretation of the Galactic radial [Fe/H] abundance profile (e.g. \citealt{Lepine2011, Genovali2014, Anders2017}), one should take into account that outer-disc ($R_{Gal}>15$ kpc) CCs are on average more than 150 Myr old, while inner-disc ($R_{Gal}<5$ kpc) CCs are less than 50 Myr old, which means that the outer-disc CCs have had three times as much time to migrate away from their birth places. We illustrate this effect in Fig. \ref{fig:gradient}, again using the fundamental-mode CCs with spectroscopic [Fe/H] measurements catalogued by \citet{Ripepi2022}. The figure shows that the increased scatter about the [Fe/H] gradient observed in outer-disc CCs \citep[see also][]{Luck2018, Kovtyukh2022}, while also due to the larger absolute distance uncertainties in that more distant Galactic area, could in part also be explained by the older age of this population, as illustrated by the arrows at the bottom of the plot (corresponding to the radial migration scales in the diffusion model of \citealt{Frankel2018}, which we use here for illustrative purposes). Taking into account these systematics, along with the effects of the Galactic bar, spirals, and warp (e.g. \citealt{Semczuk2023, Dehnen2023, Cabrera-Gadea2024, Poggio2024, Drimmel2024}), will help future modeling of the detailed chemodynamical evolution of the Galactic disc. 

\section*{Acknowledgments}

We thank Bertrand Lemasle, Annie Robin, Roger Mor, Jesper Storm, Irene Tapial, Teresa Antoja, and Mercè Romero for initial collaborative work on Cepheids during the past years. We also thank Dorota Skowron and the anonymous referees for their comments, which greatly improved the quality of this paper.

The European Union grants RTI2018-095076-B-C21, PID2021-122842OB-C21, PID2021-125451NA-I00 and CNS2022-135232, and the Institute of Cosmos Sciences University of Barcelona (ICCUB, Unidad de Excelencia ’Mar\'{\i}a de Maeztu’) through grant CEX2019-000918-M. FA acknowledges financial support from MCIN/AEI/10.13039/501100011033 through grants IJC2019-04862-I and RYC2021-031638-I (the latter co-funded by the European Union NextGenerationEU / PRTR). CP acknowledges funding from the European Union's Horizon 2020 research and innovation programme under the Marie Skłodowska-Curie grant agreement No. 101072454 (MWGaiaDN). The preparation of this work has made use of NASA's Astrophysics Data System Bibliographic Services and the open-source Python packages \texttt{astropy} \citep{Astropy2018}, and \texttt{numpy} \citep{VanderWalt2011}. The figures in this paper were produced with \texttt{matplotlib} \citep{Hunter2007}. 

\section*{Supporting information}
The data and code producing the figures in this paper are available at {\tt \url{https://github.com/fjaellet/cepheids_age_distribution}}.


\bibliography{main_cepheids}

\end{document}